\relax
\documentstyle[prl,aps,epsf,twocolumn,graphics,rotate]{revtex}
\def\be{\begin{equation}}
\def\ee{\end{equation}}
\def\bea{\begin{eqnarray}}
\def\eea{\end{eqnarray}}

\begin {document}

\draft
\title{Statistical properties of localisation--delocalisation
transition in one dimension}
\author{M. Steiner$^1$, Yang Chen$^1$, M. Fabrizio$^2$, and
Alexander O. Gogolin$^1$,
}
\address{
$^1$Department of Mathematics, Imperial College, 180 Queen's Gate,
London SW7 2BZ, United Kingdom\\
$^2$International School for Advanced Studies 
Via Beirut 4, 34014 Trieste, Italy 
and Istituto Nazionale della Fisica della Materia INFM, Italy
}
\date{Draft: \today}
\maketitle
\begin{abstract}
We study a  one-dimensional model of disordered electrons 
(also relevant for random spin chains), which exhibits
a delocalisation transition at half-filling.          
Exact probability distribution functions for the 
Wigner time and transmission coefficient are calculated.
We identify and distinguish those 
features of probability densities that are due to 
rare, trapping configurations of the random potential
from those which are due 
to the proximity to the delocalisation transition.
%A long-time, log-normal magnetisation relaxation law is predicted
%for spin systems. 
\end{abstract}
\pacs{\rm PACS No:}

The Anderson transition in dimensions $D<3$ has recently 
attracted a renewed interest in relation to such systems as
random antiferromagnetic spin chains\cite{SC} and
high mobility Si MOSFET's\cite{2D}.
%experimental evidence of a low-temperature saturation 
%of the dephasing time in quasi 
%1D quantum wires\cite{Webb}, 
Experiments on 
both systems could not be accounted for in the standard  
scaling theory of localisation\cite{gang}. 
%In this respect, it might be useful 
%to better understand some simple model which does not fit in with the 
%scaling theory of localisation. 

The simplest disordered model known to exhibit metallic behaviour, 
in spite of being one-dimensional, is the random-hopping model,
\begin{equation}
H_{rh} = \sum_n t_n ( c^\dagger_n c^{\phantom{\dagger}}_{n+1}
+ H.c. ),
\label{random-hop}
\end{equation}
where $t_n>0$ are random variables, with $n$-independent average, 
$\langle t_n \rangle = t$, and $c_n$ annihilates a spinless 
fermion at site $n$. 
This 1D model has a single delocalised state at the middle of the band,
$\epsilon=0$. 
%Furthermore, when the chemical potential lies on this extended state, 
% WE SAY THIS LATER!
%corresponding to a half-filled lattice, the conductivity is finite. 
%thus being insulating if $E_F\not = E_c$ and metallic exactly at 
%$E_F=E_c=0$ (half-filled band)\cite{GM}.   
This is an interesting  
example of the failure of the scaling theory of localisation. 
Moreover, this model has many common features with a wide class of 
random spin chains; such as the spin--1/2 random Heisenberg chain
$
H = \sum_n J_n \vec{S}_n \cdot \vec{S}_{n+1},
$
where $J_n>0$ are randomly distributed.
(The XX version of the latter model is, in fact, exactly 
equivalent to (\ref{random-hop}), upon the Jordan-Wigner 
transformation.) 
%This is a further reason for studying 
%the model (\ref{random-hop}), since there are evidences that 
%these random spin models play an important role in determining some 
%peculiar properties of doped spin-Peierls and spin-ladder 
%compounds, which have recently been the subject of a detailed experimental 
%investigation\cite{SC}.       

For the random hopping model, 
a great deal is known about such self-averaging quantities as
the total density of states.
More recently, 
some of the correlation functions have also been 
calculated. 
However, the behaviour of probability distributions
in the proximity of the delocalisation transition
is virtually unexplored.
It is the purpose of this letter to address this issue. 
%The analysis of the probability distributions sheds
%some light onto such long-standing questions of the localisation
%theory as the relation of the log-normal tails to the
%delocalisation phenomenon.
%As a bonus, we also draw consequences for 
%relaxation processes in random spin systems. 

In the continuum limit, model (\ref{random-hop}) 
becomes what is known as the random-mass Dirac model
(see e.g. \cite{SFG,BF}):
\be
{\cal H}=-i [ R^\dagger\partial_x
R^{\phantom{\dagger}}- L^\dagger\partial_x
L^{\phantom{\dagger}}]
-im(x)[ R^\dagger
L^{\phantom{\dagger}}- L^\dagger
R^{\phantom{\dagger}}]
\label{RMD}
\ee
where $R$ and $L$ are the chiral components of the 
electron field operator. 
The derivation of the continuum limit 
assumes weak disorder such that
$t_n= t + \delta t_n$ (the Fermi velocity, $v_F$, 
associated with $t$ is set to 1). 
It is the staggered component of the
random hopping, 
$\delta t_n \rightarrow (-1)^n m(x)$,
which enters into the continuum theory. 
In field-theoretic language, this corresponds to a random mass. 
%The exact equivalence 
%with (\ref{random-hop}) holds only if the average $\langle m(x) \rangle=0$. 
%However, for the sake of completeness, we will also consider the case of a 
%finite $\langle m(x)\rangle = m_0$, which causes the disappearance of the 
%mobility edge leading to an insulating state at any density. 

It was found by Dyson in 1953 \cite{Dyson},
that the average electron density of states for a
model equivalent to (\ref{random-hop}) 
diverges at the middle of the band:
$\rho(\epsilon)\sim 1/(\epsilon |\ln \epsilon |^3)$.
By the Thouless relation\cite{Thouless},
such a density of states implies a divergent 
localisation length $\lambda_\epsilon \sim |\ln \epsilon|$. 
The criticality of the model at half-filling 
%where the Fermi energy 
%crosses the mobility edge, 
was established by Gogolin and Mel'nikov\cite{GM}.  
%They obtained asymptotic forms of two-particle 
%correlation functions which turned out to obey a power law
%decay ($\sim |x|^{-3/2}$) at large distances and,
In particular, they found that model (\ref{RMD})
has a finite conductivity in contrast with the Mott law
in the standard localised regime.
Calculations of Ref.\cite{George} first indicated that it is 
not the Thouless length $\lambda_\epsilon$, but rather the
length $l_\epsilon \sim \ln^2\epsilon$,
which is likely to govern the correlation functions. 
%We shall call $l_\epsilon$ the correlation length;
%obviously, it is still divergent at half-filling.
The role of the length $l_\epsilon$ was
later clarified by means of the real-space renormalisation group
method by Fisher \cite{Fisher}. 
The cross over for the electron Green function
between the spatial regions $x\ll l_\epsilon$ (critical
regime) and $l_\epsilon \ll x$ (localised regime) was discussed
in recent publications \cite{SFG} and \cite{BF}
by means of the Berezinskii
technique and the Efetov supersymmetry method, respectively. 

It seems natural to undertake the next step and investigate
the probability distribution functions for the above systems.
To this end, we adopt a simple mathematical
technique, based on the recent observation by Shelton
and Tsvelik\cite{ST} that, at $\epsilon=0$, the random
mass Dirac model can be formulated as a  one-particle 
quantum mechanical problem.
Indeed, if we introduce the combinations 
$\chi_\pm=(R \mp L)/\sqrt{2}$ of the chiral components
of the electron field, then the
Dirac equation becomes
\[
[\partial_x \mp m(x)]\chi_\pm=i\epsilon \chi_\mp \;.
\]
This equation decouples at $\epsilon=0$,
thus admitting the (unnormalised) solutions of the form
\be
\chi_\pm (x) \sim e^{\pm V(x)}. \;\;\;
V(x)= \int_0^x dy m(y)
\label{chisol}
\ee

As it is customary in the literature, we assume that the 
random mass $m(x)$ is 
$\delta$-correlated in the real space and Gaussian:
%\cite{CDM}
%\be
%P_{m_0}[m(x)]\sim \exp \left\{-\frac{1}{2g}\int dx [
%m(x)+m_0\right]^2\right\}
%\label{Gauss}
%\ee
$P_{m_0}[m(x)]\sim \exp [-\left(m(x)-m_0\right)^2/2g]$.
Here $g$ characterises the disorder strength and
the mass has a mean value, $m_0$. 
%(Owing to the fact that $m(x)$ corresponds to the staggered 
%component of the exchange coupling, having a finite mean mass
%is equivalent to introducing a constant dimerisation field
%for the spin systems.)
%The integral $V(x)$ of the Gaussian variable $m(x)$ has a
%statistics of a random walk. 
Thus, solution (\ref{chisol}) for the wave-functions is
nothing but an exponential of a random walk with
a drift term, $m_0 x$.

Wave-function (\ref{chisol}) allows us to investigate the distribution 
of several physical quantities in a relatively simple manner. 
We start by considering the so-called relaxation time, which was 
introduced in the context of the scattering theory by Wigner \cite{Wigner},
hence also known as the Wigner time. 
Physically this is the time 
spent by a wave packet inside a scattering region, and it can be formally 
defined as the momentum derivative of the scattering phase shift. 
Let the effect of the disorder be confined to the segment $[0,L]$, which 
is the scattering region in our problem. 
For the sake of concreteness, we assume that the electrons
can not leave the sample on the left-hand-side ($x<0$),
so that they are scattered off the segment $[0,L]$ on the
right. We thus impose a vanishing boundary condition 
on the left, $\chi_-(0)=0$ (this condition corresponds
to suppressing a site in the lattice formulation, (\ref{random-hop})).
The boundary condition on the right is simply the continuity
of the Dirac wave-function at $x=L$.  
%($s$-wave 
%scattering phase in Wigner's three dimensional formulation).
In this case, the calculation of the 
Wigner time involves the scattering phase
picked up by the incident left-moving wave in the process
of scattering from the disordered segment, 
%(and converted
%to a right-moving wave in the course of this process),
so that ($x>L$):
\be
\tau_\epsilon=d\theta/d\epsilon,\;\;\;
\theta(\epsilon)=-i\ln\left[e^{-2i\epsilon x}
R(x)/L(x)\right].
\label{taudef}
\ee

We are mainly interested in the relaxation
time at half-filling, i.e.  
%i.e. in the properties of 
$\tau_0$.
In order to find this, we perturb the Dirac
equation in $\epsilon$ around the $\epsilon=0$
solution (\ref{chisol}).
%, demand the boundary conditions 
%at $x=0,L$ be satisfied for the perturbed wave function,
%and use the definition (\ref{taudef}). 
Performing this simple calculation we found the following
exact expression for the Wigner time 
as a functional of the disorder\cite{CT}  
\be
\tau_0[V]= 2\int_0^L dx \left\{ e^{2[V(x)-V(L)]}-1\right\}
\label{tauV}
\ee
%An expression of this type for the Wigner time
%was recently found in Ref.\cite{CT} by a different method. 
In accordance with \cite{Wigner}, the relaxation time 
is related to the total charge, 
$Q_0=\int_0^L dx (|R_0|^2+|L_0|^2)$, 
and therefore $\tau_0=Q_0-2L$.

As such, expression (\ref{tauV}) does not supply much information.
Indeed, the physical content of the problem is revealed by
the probability density of the relaxation time
${\cal P}[\tau_0] =\langle \delta\left( \tau_0-\tau_0[V] \right)
\rangle$.
%In the spirit of \cite{ST},
We observe that the Laplace transform of this density,
$\hat{P}[\lambda]=\int_0^\infty d\tau_0 e^{-\lambda \tau_0}
{\cal P}[\tau_0]$, can be represented as a path integral
\be
\hat{P}[\lambda]\sim \int DV(x) P_{m_0}[V]
e^{-\lambda \tau_0[V]}
\label{pathint}
\ee
In the above formula, the integration is taken along all
paths starting at $V(x=0)$ and ending at $V(x=L)=V_0$.
By construction, see (\ref{chisol}), the starting point
is $V(0)=0$, while the ending point $V_0$ is arbitrary,
so the path integral in (\ref{pathint}) involves
an additional integration over all possible end points $V_0$.
Fixing the normalisation, $\hat{P}[0]=1$,
and performing a convenient shift of the integration
variables, we ultimately obtain 
\be
\hat{P}[\lambda]=\int_{-\infty}^{\infty} dV_0 
K_\lambda (V_0,0; L)
\label{lambdaPK}
\ee
%for the Laplace transformed distribution function.
(For the time being we have set $m_0=0$, so as to be at
the criticality.)
Here $K_\lambda (V_1, V_2; x)$ is the imaginary-time 
propagator for a quantum mechanical particle characterised
by the action
\be
S_\lambda [V]=\int_0^L dx \left[ \frac{1}{2g} (\partial_x V)^2
+2\lambda (e^{2V}- 1) \right]\;.
\label{action}
\ee

The system described by this action
is known as Liouville quantum mechanics \cite{Liouv}.
The Schr\"{o}dinger equation, corresponding to the 
action (\ref{action}), has the following (normalised) 
solution \cite{Liouv}
\be
\psi_\gamma(V)= \sqrt{2\gamma \sinh \pi \gamma/\pi^2}
K_{i \gamma}\left(2\sqrt{\lambda/g} e^V\right)
\label{psi}
\ee
with the energy $-2\lambda+g\gamma^2/2$ ($K_{i\gamma}$
is the MacDonald function).
Constructing the propagator in a standard manner by 
making use of the complete set (\ref{psi}), integrating
over $V_0$, and
%in the form:
%\bea
%\hat{P}[\lambda] &=&
%\frac{2}{\pi}\int_0^\infty d\gamma \cosh(\pi\gamma/2)
%K_{i \gamma}\left(2\sqrt{\lambda/g} \right)
%e^{2\lambda L- g\gamma^2L/2}
%\nonumber\\
%~&=& \sqrt{\frac{2}{\pi gL}}
%e^{2\lambda L+ \pi^2/8 gL}
%\int_0^\infty dt \cos \left( \frac{\pi t}{2 gL}\right)
%e^{-2\sqrt{\lambda/g} \cosh t - t^2/2gL}
%\label{Plres}
%\eea
%The last line (manifestly normalised) follows from using an
%integral representation for the MacDonald function.
%Finally, 
performing the inverse Laplace transform, 
we obtain the exact Wigner time probability density 
\bea
{\cal P}[\tau_0]&=&
\frac{2^{1/2} e^{\pi^2/8gL}}{\pi g L^{1/2} Q_0^{3/2}}
\int_0^\infty dt \cosh t \cos \left( \frac{\pi t}{2 gL}\right)
\nonumber\\
~&~&\exp\left[-\cosh^2 t/(g Q_0) -t^2/(2gL)\right]\;.
\label{Ptau-fin}
\eea
It is convenient to represent the result in terms
of the positively definite quantity (total charge) 
$Q_0= \tau_0 +2 L$. According to the general theory of 
Ref.\cite{Wigner}, the delay time can be negative (for
repulsive potentials) but there is a lower bound,
which is $-2L$ in our case.

We note the following interesting limiting cases 
of the formula
(\ref{Ptau-fin}).\\
$\bullet$ Consider first the probability of
long time delays, $\tau_0\to \infty$ (and therefore $Q_0
\to \infty$). 
The $t$-integral in (\ref{Ptau-fin}) is convergent for all $Q_0$.
It is therefore tempting to expand the exponential in the 
integrand in powers of $1/Q_0$. 
It is easy to check, however,
that all the coefficients of such an expansion identically
vanish. 
It follows that (\ref{Ptau-fin}) has an essential
singularity at $\tau_0=\infty$.
The nature of this singularity can be determined by first
neglecting the $1/Q_0$ term in the exponential 
and then simulating its effect by cutting off the $t$-integration
at large times $t_0\sim (1/2) \ln (g \tau_0)$, when the
term becomes of the order of unity. 
Then, within the leading logarithmic accuracy, we obtain:
\be
{\cal P}[\tau_0] \sim \exp \left[- \ln^2(g\tau_0)/8gL
\right]
\label{ln}
\ee
This formula is valid when the factor in the exponential
is large, i.e. $\ln^2 \tau_0 \gg L$.
As a simple application to random spin chains, we consider 
long-time relaxation of the magnetisation $M(t)$ 
inside a finite segment of length $L$. 
Due to (\ref{ln}) we find that 
$M(t) \sim \exp{[-\ln^2(t)/8gL]}$: a very slow 
decay.\\ 
$\bullet$ Since the applicability 
of (\ref{ln}) involves the system size 
$L$, the probability distribution for large samples
ought to be different. 
Indeed, for a fixed $Q_0$, we obtain:
\be
{\cal P}[\tau_0]|_{L\to \infty} \simeq
\frac{1}{\sqrt{2\pi g L} Q_0}e^{-1/g Q_0}
\label{Pcritical}
\ee

The next quantity of interest is the transmission 
coefficient at $\epsilon=0$, which is also proportional to the 
Landauer conductance  at half-filling. 
In order to have a finite transmission coefficient $T$, 
let us open our sample 
on the left-hand-side, so that the electrons can now leave it
at $x=0$ 
(where the boundary condition thus becomes the same as at $x=L$).
Upon matching the wave-functions in the usual way, one finds
\be
T[V]= 1/\cosh^2 V(L)\;.
\label{Tdef}
\ee
This is a simple formula: the transmission 
coefficient in not a functional of the entire random walk
trajectory (as the Wigner time is) but only a function
of the end point. 
Therefore the probability distribution $P[T]$
can be found in an elementary way without using 
the Liouville mechanics.
We obtain
\be
P[T]= 
%\sqrt{\frac{2}{\pi g L}}\int_0^\infty dV 
%{\rm e}^{-\frac{V^2}{2gL}}\delta\left(T - \frac{1}{\cosh^2 V}\right)=
\frac{\sqrt{1/2\pi gL}}{T\sqrt{1-T}}
\exp\left\{-\frac{1}{2gL}\left[\cosh^{-1}(1/\sqrt{T})\right]^2
\right\}
\label{PT-fin}
\ee
where $0\leq T \leq 1$.
This is an intriguing distribution function, plotted in 
Fig.1. It has the following properties:\\
$\bullet$ The probability of a small transmission is:
\be
P[T] \sim \exp \left[ -
\ln^2(1/T)/8gL\right]
\label{lnT}
\ee
$\bullet$ The function $P[T]$ has a low-transmission peak
at $T_0 \sim e^{-4gL}$.\\ 
$\bullet$ There is an integrable divergence close to the
perfect transmission ($T=1$):
\be
P[T]\simeq \sqrt{8/\pi gL(1-T)} 
\label{perfect}
\ee
$\bullet$ The mean transmission coefficient is given by
\be
\langle T\rangle \simeq \sqrt{2/\pi g L}
\label{meanT}
\ee
This formula is asymptotically exact as $L\to \infty$
(we also verified this result by an independent 
calculation using the Berezinskii technique \cite{unpublished}).
This result is surprising. 
Since $T\sim \sigma/L$, $\sigma$ being the dc-conductivity,
the $1/\sqrt{L}$ behaviour of the 
average transmission coefficient suggests that the system is 
even more metallic than what one would expect in
an hypothetical case of a weakly disordered 1D metal 
(hypothetical because any weak disorder is supposed 
to lead to localisation).
It is worth noting that the transmission coefficient is not a good 
scaling variable in the limit $L\to \infty$. 
So, from Eq.(\ref{PT-fin}), 
one finds that the variance of the transmission coefficient 
$\delta^2 = \langle (T-\langle T \rangle)^2\rangle \sim 1/\sqrt{L}$, implying 
that the width of the distribution, normalised to the mean transmission 
$\delta/\langle T \rangle \sim L^{\frac{1}{4}}$, 
diverges when $L\to \infty$.
On the other hand, the logarithm of the transmission coefficient is a good 
scaling variable in the above sense, with average 
\[
\langle \log\left(\frac{1}{T}\right) \rangle = \alpha\sqrt{gL}, 
\]
$\alpha$ being a positive numerical coefficient. 
This conclusion was already reached in 1980 by Anderson {\sl et al.} in 
studying a generic 
1D disordered system\cite{PW}. 
They introduced  the {\sl scale conductance},
$
T_{typ} = {\rm e}^{\langle \log T \rangle},
$
instead of the average conductance. 
In our model we reach a similar conclusion.  
However, unlike the standard case, we find that 
$
T_{typ}= {\rm e}^{-\alpha \sqrt{g L}}
$
behaves quite differently from the average transmission, and both 
are different from the peak in the probability distribution 
$T_0 \sim e^{-4gL}$.

There is a difference between our $1/\sqrt{L}$
result for the Landauer conductance and the 
finite conductivity found for an infinite system in Ref.\cite{GM}.
A possible physical explanation is that this difference is due to resonant
scattering processes, which enhance the probability of 
near-perfect transmission in finite samples but are absent in
infinite systems (this is characteristic for a critical 
system; otherwise, the $T\to 1$ divergence of $P[T]$
is exponentially suppressed in the sample length). 
However, this question requires further studies.

Both calculations for ${\cal P}[\tau_0]$ and $P[T]$
can be generalised to the off-critical case when $m_0$ is non-zero. 
The details will be given in an extended article
\cite{unpublished}. 
Here we only quote the results. 
Formulae (\ref{ln}) and (\ref{lnT}) are not affected.
The power law in the denominator of (\ref{Pcritical})
changes to $1/Q_0^{1+m_0/g}$. 
Both formulae (\ref{perfect})
and (\ref{meanT}) acquire a suppressing factor, 
exponentially small in the parameter $m_0L$.

%Let us now discuss our results
%on the distribution functions.
\begin{figure}[]
\begin{center}
\leavevmode
\epsfxsize=7cm \epsffile{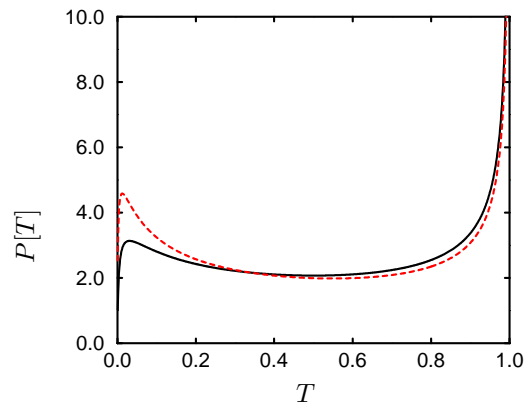}
\begin{picture}(10,10)(0,0)
\setlength{\unitlength}{5mm}
\put(-7,0.3){$T$}
\put(-14.5,4){\rotate{$P[T]$}}
\end{picture}
\caption{Distribution of the transmission coefficient 
with $gL=1.25$ and average mass zero (solid line) and
$m_0 L = 0.5$ (dashed line).}
\label{prob_t}
\end{center}
\end{figure}
Notice that the probability of
long time delays, (\ref{ln}), as well as the probability
of a small transmission, (\ref{lnT}),
follows the so-called log-normal law.
To our knowledge, the log-normal tails of the
distribution functions in one-dimensional disordered systems
were first obtained in Ref.\cite{PW} by means of a scaling
argument. 
Their existence was rigorously established by Mel'nikov
via the Bereziskii technique\cite{Mel}. 
In $2+\epsilon$ dimensions, 
the log-normal tails were found in Ref.\cite{AKL},
which was thought to be a signature the Anderson transition.
The discussion of the log-normal tails  
was recently revived by 
Muzykantskii and Khmelnitskii\cite{MK} (see
also \cite{EF}) who
gave a simple derivation 
based on a specific saddle-point approximation 
in the supersymmetric $\sigma$-model. 
(For a recent collection of results on the log-normal
distributions see \cite{Mirlin}.)

We have shown that the random
mass Dirac model does posses log-normal tails at the
criticality. 
Moreover, these tails are unaffected when
one moves away from the criticality.
This is consistent with the interpretations \cite{MK,Mirlin}
that these tails are due to the so-called `anomalously
localised' electronic  states, which occur in rare, trapping 
disorder configurations. 
Indeed our result (\ref{ln}) can be understood
in terms of the `optimal fluctuation' concept,
discussed in this context in \cite{MK}.
The optimal fluctuation in our case corresponds to
having a constant mass $m_0$ within the sample.
This would accumulate charge  
$Q_0 \propto \exp(2m_0 L)$.
The probability to have such a potential, and 
therefore such a charge is $\sim \exp(-m_0^2 L/2g)
\sim \exp(-\ln^2 Q_0/8 gL)$.

So, does it follow that the proximity to the delocalisation
transition (criticality) plays no role? 
To clarify this point we first notice that the log-normal tails
are only present in a finite system. 
%(The opposite conjecture of 
%Ref.\cite{EF} is therefore false for our model. 
(These tails do not follow from and are not directly related to
the so-called multi-critical exponents appearing in the wave-functions' 
statistics for the infinite system \cite{BF,ST}.)
The Wigner time distribution function takes a different form 
in the limit of a large system, (\ref{Pcritical}).
This can be interpreted as a `limiting' 
(`equilibrium') distribution
in terms of the Fokker-Planck equation approach \cite{CT}
and it reveals no trace of the log-normal behaviour.
It is this distribution which bears the signature of the 
criticality: it ceases to be normalisable for $m_0=0$
thus requiring a long-time cut off. 
Similarly, the log-normal tail appears at low$-T$ in the
transmission coefficient distribution function, while
the criticality shows up near the perfect transmission. 
Indeed, the divergence of $P[T\to 1]$ is always in
place (due to resonant scattering processes) but it is
exponentially suppressed in the sample length, unless  
the system is critical. 
As a result, at the criticality, the mean transmission coefficient
is not any more exponentially small  but is given by
the power-law formula (\ref{meanT}).
(Notice, though, that the peak of $P[T]$ 
still exists at the criticality.)
It follows that the disorder configurations 
leading to the log-normal tails and to the delocalisation
phenomenon act independently. 
%(at least in our model). 
They affect different 
domains of the parameter space of the problem and show up
in the distinct limits for the probability distribution functions.

We are thankful to A.M. Tsvelik, D.E. Khmelnitskii,
and B.A. Muzykantskii for interesting conversations.
M.F. acknowledges 
support by INFM, Project HTSC.

\end{document}